\documentclass[useAMS,usenatbib,psfig]{mn2e}
\usepackage{amssymb}
\usepackage{psfig}
\usepackage{times}

\title[Rapidly spinning massive black holes in AGN]
{Rapidly spinning massive black holes in active galactic nuclei:
evidence from the black hole mass function}
\author[X. Cao \& F. Li]
{ Xinwu Cao$^{1}
\thanks{E-mail: cxw@shao.ac.cn}$,
Fan Li$^{1,2}$\\
1. Shanghai Astronomical Observatory, Chinese Academy of Sciences,
80 Nandan Road, Shanghai, 200030, China\\
2. Graduate School of the Chinese Academy of Sciences, Beijing
100039, China}

\date{Accepted 2008 August 1.  Received 2008 June 16; in original form 2008 May 5}

\pagerange{\pageref{firstpage}--\pageref{lastpage}} \pubyear{}

\begin{document}

\maketitle \label{firstpage}

\begin{abstract}
The comparison of the black hole mass function (BHMF) of active
galactic nuclei (AGN) relics with the measured mass function of the
massive black holes in galaxies provides strong evidence for the
growth of massive black holes being dominated by mass accretion. We
derive the Eddington ratio distributions as functions of black hole
mass and redshift from a large AGN sample with measured Eddington
ratios given by Kollmeier et al. We find that, even at the low mass
end, most black holes are accreting at Eddington ratio
$\lambda\sim0.2$, which implies that the objects accreting at
extremely high rates should be rare or such phases are very short.
Using the derived Eddington ratios, we explore the cosmological
evolution of massive black holes with an AGN bolometric luminosity
function (LF). It is found that the resulted BHMF of AGN relics is
unable to match the measured local BHMF of galaxies for any value of
(constant) radiative efficiency $\eta_{\rm rad}$. Motivated by
Volonteri, Sikora \& Lasota's study on the spin evolution of massive
black holes, we assume the radiative efficiency to be dependent on
black hole mass, i.e., $\eta_{\rm rad}$ is low for $M_{\rm
bh}<10^8{\rm M}_\odot$ and it increases with black hole mass for
$M_{\rm bh}\ge 10^8{\rm M}_\odot$. We find that the BHMF of AGN
relics can roughly reproduce the local BHMF of galaxies if
$\eta_{\rm rad}\simeq0.08$ for $M_{\rm bh}<10^8{\rm M}_\odot$ and it
increases to $\ga 0.18$ for $M_{\rm bh}\ga10^9{\rm M}_\odot$, which
implies that most massive black holes ($\ga 10^9{\rm M}_\odot$) are
spinning very rapidly.
\end{abstract}

\begin{keywords}
(galaxies:) quasars: general---accretion, accretion discs---black
hole physics---galaxies: evolution
\end{keywords}

\section{Introduction}

It is believed that quasars are powered by accretion on to massive
black holes, and the growth of the massive black holes could be
governed by mass accretion in quasars. The massive black holes (AGN
relics) should be present in the centres of galaxies. Thus, the
luminosity functions (LF) of active galactic nuclei (AGN) provide
important clues on the growth of massive black holes. It is indeed
found that most nearby galaxies contain massive black holes at their
centres, and a tight correlation is revealed between central massive
black hole mass and the velocity dispersion of the galaxy
\citep{fm00,g00}. The black hole mass is also found to be tightly
correlated with the luminosity of the spheroid component of its host
galaxy \citep*[e.g.,][]{m98,mh03}. These correlations of the black
hole mass with velocity dispersion/host galaxy luminosity were used
to derive the mass functions of the central massive black holes in
galaxies \citep*[e.g.,][]{yt02,m04,t06,g07}. On the other hand, the
black hole mass function (BHMF) of AGN relics can also be calculated
by integrating the continuity equation of massive black hole number
density on the assumption of the growth of massive black holes being
dominated by mass accretion, in which the activity of massive black
holes is described by a LF of AGN
\citep*[e.g.,][]{c71,soltan82,ct92,sb92,m04,s04,t06}. Such
calculations on the cosmological evolution of massive black holes
were usually carried out by adopting two free parameters: the
radiative efficiency $\eta_{\rm rad}$ and the Eddington ratio
$\lambda$ for AGN. The derived BHMF of AGN relics in this way is
required to match that estimated either from velocity dispersion or
the luminosity of the spheroid of its host galaxy, which always
leads to $\lambda\sim 1$, i.e., almost all AGN are required to be
accreting close to the Eddington limit
\citep*[e.g.,][]{yt02,m04,s04,t06}.

In the last decade, several approaches for measuring the masses of
the central black holes in AGN have been developed, in which the
reverberation mapping may be the most effective one \citep{p93,k00}.
Using the tight correlation between the size of the broad-line
region and the optical luminosity established with the reverberation
mapping method for a sample of AGN, the black hole masses of AGN can
be easily estimated from their optical luminosity and width of broad
emission line. The Eddington ratios for thousands of AGN were
estimated with the analyses of the Sloan Digital Sky Survey (SDSS)
by \citet{md04}, which indicate that the mean Eddington ratio
$L_{\rm bol}/L_{\rm Edd}\simeq 0.1$ at $z\sim 0.2$ to $\simeq0.4$ at
$z\sim 2$. \citet{w04} also derived the Eddington ratio distribution
for a sample of $\sim$~500 AGN with redshifts $0\la z\la 5$. As
pointed by \citet{k06}, both these derived Eddington ratios are
heavily weighted towards high-luminosity objects due to the limited
sensitivity of SDSS. \citet{k06} estimated the Eddington ratios of
AGN discovered in the AGN and Galaxy Evolution Survey (AGES), which
is more sensitive than the SDSS \citep{k04}. The derived Eddington
ratio distribution at {\it fixed luminosity} is well described by a
single lognormal distribution peaked at $\sim 0.25$ independent of
redshift and luminosity \citep*[see][for the details]{k06}.

In this work, the Eddington ratio distribution at {\it fixed
luminosity} given by \citet{k06} is converted to that at {\rm fixed
black hole mass} by using an AGN LF. We integrate the continuity
equation for black hole number density adopting the derived
Eddington ratio distributions of AGN to calculate the BHMF of AGN
relics at different redshifts $z$, which is different from a free
parameter $\lambda$ adopted in most previous works. The resultant
BHMF of AGN relics is constrained by those estimated from the galaxy
LFs \citep{m04,t06}. The conventional cosmological parameters
$\Omega_{\rm M}=0.3$, $\Omega_{\Lambda}=0.7$, and $H_0=70~ {\rm
km~s^{-1}~Mpc^{-1}}$ have been adopted in this work.

\section{The Eddingtion ratio distribution of active galactic
nuclei}

The Eddington ratio distribution of AGN for given bolometric
luminosity can be approximated as a log-normal distribution:
\begin{equation}
\xi({l})={\frac {1}{\sqrt{2\pi}\sigma}}\exp\left[-{\frac {({
l}-\mu)^2}{2\sigma^2}} \right], \label{bolldist}
\end{equation}
where ${l}=\log\lambda$, $\lambda=L_{\rm bol}/L_{\rm Edd}$,
$\mu\simeq \log 0.25$ and $\sigma\simeq 0.3$ \citep*[see][for the
details]{k06}. We can derive the BHMF of AGN from the bolometric LF
$\Phi(z,L_{\rm bol})$:
\begin{equation}
N_{\rm AGN}(z,M_{\rm bh})=\int \Phi(z,L_{\rm bol}) {\frac {{\rm
d}\log L_{\rm bol}} {{\rm d}\log M_{\rm bh}} }\xi({ l}){\rm d}{ l},
\label{nagn}
\end{equation}
where $\log L_{\rm bol}={ l}+\log L_{\rm Edd}$, $L_{\rm
Edd}=1.251\times10^{38} M_{\rm bh}{\rm ~erg~s}^{-1}$, and  $M_{\rm
bh}$ is in units of solar mass. Using the bolometric LF
$\Phi(z,L_{\rm bol})$ of AGN, the Eddington ratio distribution for
given black hole mass $M_{\rm bh}$ can be calculated with
\begin{equation}
\chi(z,M_{\rm bh},{ l})={\frac {\Phi(z,L_{\rm bol})\xi({ l})}{N_{\rm
AGN}(z,M_{\rm bh})}}, \label{mbhldist}
\end{equation}
where $L_{\rm bol}=10^l L_{\rm Edd}$, and the BHMF of AGN, $N_{\rm
AGN}(z,M_{\rm bh})$, is available with Equation (\ref{nagn}). The
mean Eddington ratio of AGN with $M_{\rm bh}$ at $z$ is
\begin{equation}
\bar{\lambda}(z,M_{\rm bh})=\int\lambda\chi(z,M_{\rm bh},{ l}){\rm
d}{l}. \label{barlambda}
\end{equation}

{  In this work, we adopt the luminosity-dependent density evolution
(LDDE) bolometric LF calculated from the rest-frame optical, soft
and hard X-ray, and near- and mid-IR bands in the redshift interval
$z=0-6$ by \citet{h07}:
\begin{displaymath}
\Phi (L_{\rm bol}, z)=\Phi (L_{\rm bol}, 0)e_{\rm d}(L_{\rm bol},
z)
\end{displaymath}
\begin{equation}
={\frac {\phi_*}{(L_{\rm bol}/L_*)^{\gamma_1}+(L_{\rm
bol}/L_*)^{\gamma_2}}}e_{\rm d}(L_{\rm bol}, z). \label{h07lf}
\end{equation}
The density function $e_{\rm d}$ is given by
\begin{equation}
e_{\rm d}(L_{\rm bol}, z)=\left\{ \begin{array}{ll}
    (1+z)^{p1},      & z \le z_{\rm c},\\
      (1+z_{\rm c})^{p1} [(1+z)/(1+z_{\rm c}(L_{\rm bol}))]^{p2}     &  z > z_{\rm c}.
      \end{array} \right. \label{e_d}
        \end{equation}
with
\begin{equation}
z_{\rm c}(L_{\rm bol})=\left\{ \begin{array}{ll}
   z_{\rm c,0} (L_{\rm bol}/L_{\rm c})^\alpha & (L_{\rm bol} \leq
L_{\rm c}),\\
z_{\rm c,0} & (L_{\rm bol} > L_{\rm c}),
      \end{array} \right. \label{z_c}
        \end{equation}
and
\begin{equation}
p1(L_{\rm bol})=p1_{46}+\beta_1[\log(L_{\rm bol}/10^{46}{\rm ergs~
s^{-1}})],
\end{equation}
\begin{equation}
p2(L_{\rm bol})=p2_{46}+\beta_2[\log(L_{\rm bol}/10^{46}{\rm ergs~
s^{-1}})].
\end{equation}
All the parameters of the LF are as follows:
$\log\phi_*=-6.20\pm0.15~{\rm Mpc^{-3}}$, $\log L_{*}({\rm
ergs~s^{-1}})=45.99\pm0.10$, $\gamma_1=0.933\pm0.045$,
$\gamma_2=2.20\pm0.14$, $\log L_{\rm c} ({\rm
ergs~s^{-1}})=46.72\pm0.05$, $z_{\rm c,0}=1.852\pm0.025$,
$\alpha=0.274\pm0.025$, $p1_{46}=5.95\pm0.23$,
$p2_{46}=-1.65\pm0.21$, $\beta_1=0.29\pm0.34$, and
$\beta_2=-0.62\pm0.17$ \citep*[see Table 4 in][]{h07}.  }

\begin{figure}
\centerline{\psfig{figure=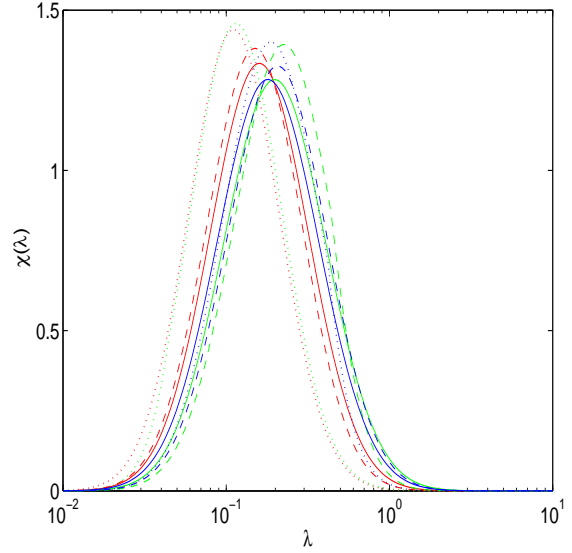,width=7.5cm,height=7.5cm}}
\caption{The Eddington ratio distributions for given black hole mass
$M_{\rm bh}$.  The red, green, and blue lines represent $z=0$, 1,
and 3, respectively. The solid, dashed, and blue lines are for
$M_{\rm bh}=10^7$, $10^8$, and $10^9~{\rm M}_\odot$, respectively. }
\label{fig1}
\end{figure}

\begin{figure}
\centerline{\psfig{figure=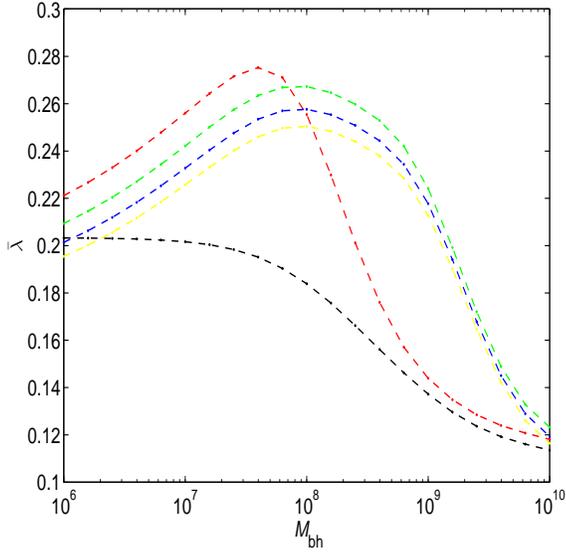,width=7.5cm,height=7.5cm}}
\caption{The mean Eddington ratios as functions of black hole mass
$M_{\rm bh}$ for the AGN at different redshifts: $z=0$(black),
1(red), 2(green), 3(blue)  and 4(yellow). } \label{fig2}
\end{figure}

In Fig. \ref{fig1}, we plot the Eddington ratio distributions
$\chi(z,M_{\rm bh},{ l})$ for fixed black hole mass derived from
that for {\it fixed luminosity} given by \citet{k06}. We find that
the derived Eddington ratio distributions are close to the lognormal
distribution, while their peaks vary with black hole mass and
redshift.  We plot the mean Eddington ratios as functions of black
hole mass in Fig. \ref{fig2}. It is found that the mean Eddington
ratios $\bar{\lambda}(z,M_{\rm bh})$ are in the range of $\sim
0.1-0.3$ as functions of black hole mass $M_{\rm bh}$ and redshift
$z$.


\section{The evolution of massive black holes}

The evolution of massive black hole number density is described by
\citep{sb92}
\begin{equation}
{\frac {\partial N(M_{\rm bh},t)}{\partial t}}+{\frac
{\partial}{\partial M_{\rm bh}}}[N(M_{\rm bh},t)<\dot{M}(M_{\rm
bh},t)>]=S(M_{\rm bh}, t), \label{mbhevol}
\end{equation}
where $N(M_{\rm bh},t)$ is the mass function of massive black holes
including both active and inactive black holes, $<\dot{M}(M_{\rm
bh},t)>$ is the mean mass accretion rate for the black holes with
$M_{\rm bh}$, {  and $S(M_{\rm bh}, t)$ describes the effect of
black hole mergers on the BHMF. The total black hole mass density
will not be altered by mergers, if the mass loss caused by the
gravitational radiation is neglected. \citet{s07} assessed the
importance of black hole mergers on the evolution of the BHMF using
a simple mathematical model that assumes constant probability
$P_{\rm merg}$ of equal mass mergers per Hubble time, similar to the
models of \citet{r98}. They found that the effect of black hole
mergers is to slightly lower the number density of small black holes
and increase the number density of massive black holes, if $P_{\rm
merg}=0.5$ is adopted \citep*[see Fig. 13 in][]{s07}. Observational
estimates of the galaxy merger rate and its mass dependence span a
substantial range \citep*[e.g.,][]{bps06,chw07,mhb08}, and $P_{\rm
merg}=0.5$ is roughly consistent with the high end of these
estimates \citep*[see][for the detailed discussion]{s07}. Thus, they
concluded that the impact of mergers on the BHMF is rather small
compared with mass accretion. } In this work, we neglect the effect
of black hole mergers, i.e., $S(M_{\rm bh}, t)\equiv 0$ in Eq.
(\ref{mbhevol}), as in most of the previous works
\citep*[e.g.,][]{yt02,m04,t06}. The mean mass accretion rate is
\begin{equation}
<\dot{M}(M_{\rm bh},t)>={\frac {\bar{\lambda}(M_{\rm bh},t)L_{\rm
Edd}}{\eta_{\rm rad} c^2}}\delta(M_{\rm bh},t), \label{meanmdot}
\end{equation}
where the duty cycle of active black holes is defined as
\begin{equation}
\delta(M_{\rm bh},t)= {\frac {N_{\rm AGN}(M_{\rm bh},t)}{N(M_{\rm
bh},t)}}.  \label{dutycycle}
\end{equation}
Substituting Eqs. (\ref{meanmdot}) and (\ref{dutycycle}) into Eq.
(\ref{mbhevol}), we can rewrite the black hole evolution equation as
\begin{equation}
{\frac {\partial N(z,M_{\rm bh})}{\partial z}}=-{\frac {{\rm
d}t}{{\rm d}z}}{\frac {\partial}{\partial M_{\rm bh}}}\left[ {\frac
{\bar{\lambda}(z,M_{\rm bh})L_{\rm Edd}{N_{\rm AGN}(z,M_{\rm
bh})}}{\eta_{\rm rad} c^2}}\right]. \label{bhmevol2}
\end{equation}

The AGN LF plays an important role in the study of the cosmological
evolution of massive black holes. The optical quasar LF was adopted
in \citet{yt02}, however, the optical quasar LF
\citep*[e.g.,][]{b00} has missed faint AGN (either intrinsic
low-luminosity or obscured AGN). The hard X-ray surveys ($\sim
2-10$keV) can trace the whole AGN population, including obscured
type II AGN. The hard X-ray LF derived by \citet{u03} was used in
some works on the cosmological evolution of massive black holes
\citep*[e.g.,][]{m04,t06,s07}. The number density of Compton-thick
AGN is still quite uncertain, which is not included in the hard
X-ray LF. The contribution of Compton-thick AGN to the black hole
evolution was taken into account by multiplying a correction factor
of 1.6 independent of the luminosity \citep*[e.g., see][]{m04,t06}.

The hard X-ray ($>20~{\rm keV}$) and the mid-IR ($5-50~\mu{\rm m}$)
bands are optimal for detection of AGN with column densities $\la
10^{24}~{\rm cm}^{-1}$ \citep*[e.g.,][]{tu05}. The observations with
the International Gamma-Ray Astrophysics Laboratory (INTEGRAL) and
the Swift Burst Alert Telescope (BAT) indicate that the fraction of
the absorbed AGN decreases with the $20-100$~keV luminosity
\citep{m05,b06}\citep*[but also see][]{wj06}, which is confirmed by
the mid-IR Spitzer observations of 25 luminous and distant quasars
\citep{m07}. \citet{m07} suggested that the fraction of the obscured
AGN to the total can be well fitted with
\begin{equation}
f_{\rm obsc}={\frac {1}{1+{\cal L}_{\rm opt}^{0.414}}},
\label{fobsc} \end{equation} where
\begin{displaymath}
{\cal L}_{\rm opt}={\frac {\lambda L_{\lambda}(5100{\rm \AA}){\rm
[erg~s^{-1}]}}{10^{45.63}}}.
\end{displaymath}
\citet{mh07} found that the fraction of type II AGN detected in the
hard X-ray band can be described by this function (Equation
\ref{fobsc}) quite well (see Fig. 3 in their paper). \citet{h07}
suggested that the fraction of Compton-thick to the total also
decreases with luminosity and it is less than $\sim 30$ per cent
based on a variety of very hard X-ray/soft gamma-ray observations on
AGN (see their paper for the detailed discussion). Motivated by the
results of these works, besides the luminosity-independent
correction for Compton-thick AGN, we tentatively employ a similar
luminosity-dependent correction as that given by \citet{m07}. { We
assume the number ratio of Compton-thick to Compton-thin AGN to be
\begin{equation}
f_{\rm CT}={\frac {0.6}{1+{\cal L}_{\rm bol}^{0.414}}},
\label{fcompth}
\end{equation}
where ${\cal L}_{\rm bol}=L_{\rm bol}/10^{46.6}$, as $L_{\rm
bol}\simeq 9\lambda L_{\lambda}(5100{\rm \AA})$ is adopted
\citep*[e.g.,][]{k00}}. {  We change the numerator in Eq.
(\ref{fobsc}) to $0.6$ here, so that $f_{\rm CT}$ reduces to $\sim
0.6$ for low-luminosity AGN,} which is the same as that in
\citet{m04}, while $f_{\rm CT}\rightarrow 0$ for luminous AGN.

\begin{figure}
\centerline{\psfig{figure=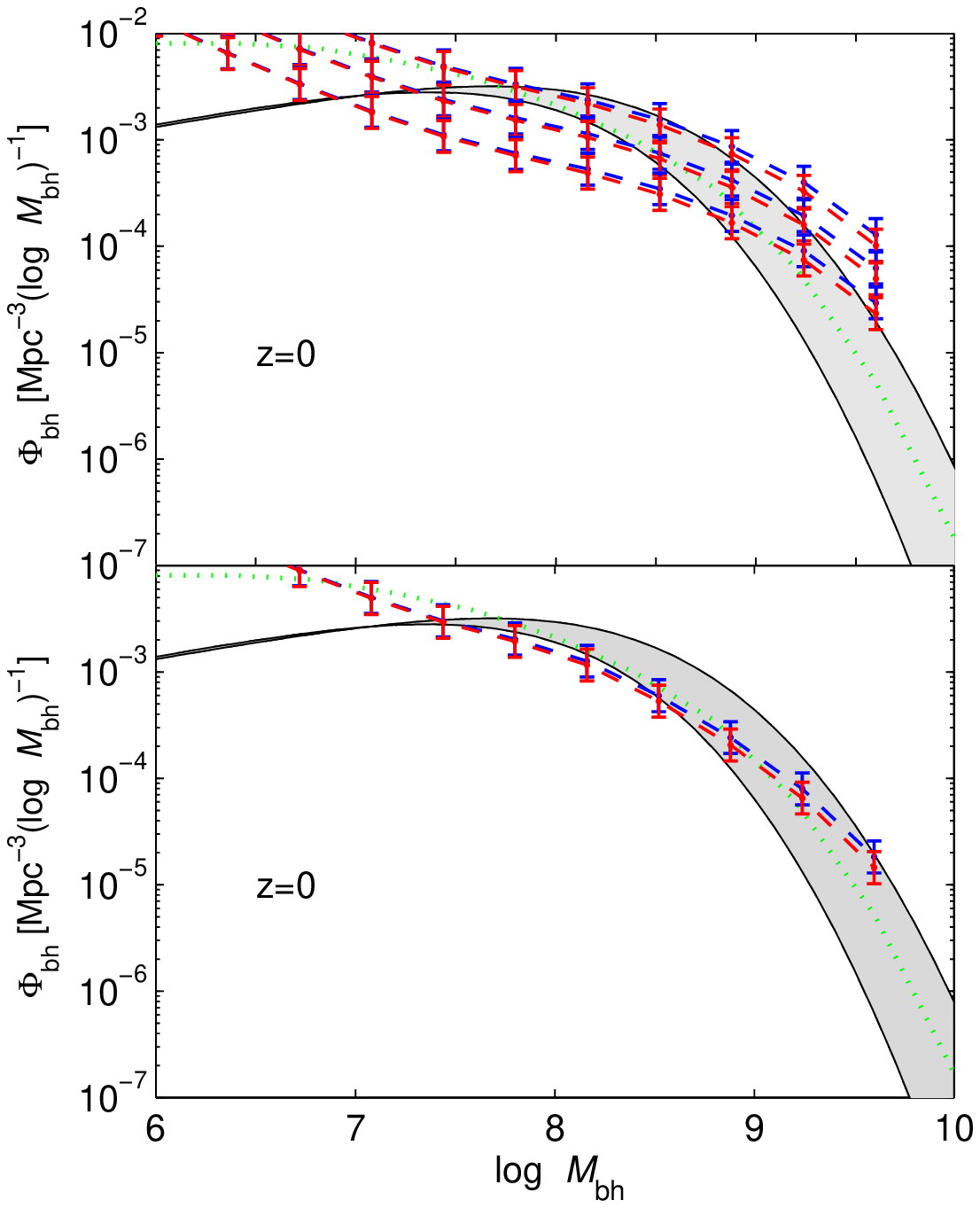,width=7.5cm,height=9.0 cm}}
\caption{The local black hole mass functions. The green dotted line
represents the local BHMF derived by \citet{m04} (see their paper
for the details). The shaded area is the BHMF transformed from the
COMBO-17 LF by using $M_{\rm bh}-L$ relation \citep*[taken
from][]{t06}. The BHMFs of AGN relics calculated from the AGN LF are
plotted in the upper panel for different radiative efficiencies
($\eta_{\rm rad}=0.05$, 0.1, and 0.2, from up to down). In the lower
panel, we calculate the BHMFs of AGN relics from the AGN LF using
$M_{\rm bh}$-dependent radiation efficiency $\eta_{\rm rad}(M_{\rm
bh})$ (see the text for the details). The blue lines are calculated
with a correction factor of 1.6 in dependent of luminosity for the
Compton-thick AGN, while the red lines represent the cases with a
luminosity-dependent correction for the Compton-thick AGN (see the
text for the details).  } \label{fig3}
\end{figure}

In most of the previous works, both the radiative efficiency of
$\eta_{\rm rad}$ and Eddington ratio $\lambda$ are free parameters,
and the comparisons between the BHMF of AGN relics and the measured
local BHMF of galaxies always require: $\eta_{\rm rad}\sim 0.1$ and
$\lambda\sim 1$ \citep*[e.g.,][]{yt02,m04,t06}. As we have derived
the mean Eddington ratio distributions as functions of black hole
mass and redshift in the last section, there is only one free
parameter $\eta_{\rm rad}$ in our calculations for the cosmological
evolution of massive black holes. The local BHMFs estimated by using
the correlation of the black hole mass with host galaxy luminosity
are adopted in this work \citep*[see,][for the details]{m04,t06}.
The continuity equation (\ref{bhmevol2}) for black hole number
density is integrated from $z=z_{\rm max}$ by using Eqs.
(\ref{nagn})-(\ref{barlambda}) and assuming the duty cycle is 0.5 at
$z_{\rm max}$. The final results are insensitive to the initial
conditions at $z_{\rm max}$. In all our calculations, $z_{\rm
max}=4$ is adopted, because the Eddington ratio distributions are
calculated from a sample of AGN with $z<4$ \citep{k06}. The resulted
BHMFs of AGN relics at low redshifts are insensitive to the value of
$z_{\rm max}$, because the fraction of local black hole mass
accreted at high redshifts can be neglected. We plot our results
with different values of $\eta_{\rm rad}$ in the upper panel of Fig.
\ref{fig3}, which indicates that the measured local BHMF cannot be
fitted with any values of $\eta_{\rm rad}$. This is due to the mean
Eddington ratios derived in this work being $\sim 0.1-0.3$, which
deviates significantly from $\lambda \sim 1$ suggested in most of
the previous works \citep*[e.g.,][]{yt02,m04}. In our calculations,
we only consider the uncertainty of the number density in the
bolometric LF given by \citet{h07}.

{  The radiative efficiency of black hole accretion is closely
related to the black hole spin. The massive black holes will be spun
up through accretion, as the black holes acquire mass and angular
momentum simultaneously through accretion. The spins of the massive
black holes may also be affected by mergers of black holes. A
rapidly rotating new black hole will be present after the merger of
two black holes, only if the binary's larger member already spins
quickly and the merger with the smaller hole, or if the binary's
mass ratio approaches unity \citep{hb03}. The comoving space density
for heavier black holes is much lower than that for smaller black
holes \citep*[e.g., see the BHMF in][]{m04}, which means that the
probability of the mergers of two black holes with similar masses is
lower for heavy  black holes. This implies that the spins of heavy
black holes are mainly regulated by accretion rather than the
mergers, and the spin parameters can reach $\sim 1$ after their
masses are doubled through accretion.
For smaller black holes in disk galaxies, a small number of minor
mergers might have happened, which are believed to be responsible
for rebuilding their host galaxy disks, and small minor accretion
episodes on to black holes are triggered by these minor mergers
\citep*[e.g.,][]{v07,kph08}. This is also supported by the
observations that single accretion events last $\sim 10^{5}$ years
in Seyfert galaxies and their total active lifetime is $10^8-10^9$
years \citep{kob06,h97,v07}. \citet{v07} studied on how the
accretion from a warped disc influences the evolution of black hole
spins and concluded that within the cosmological framework, one
indeed expects most supermassive black holes in elliptical galaxies
to have on average higher spin than black holes in spiral galaxies.
The random small accretion episodes (e.g., tidally disrupted stars,
accretion of molecular clouds) might have played a more important
role on the spin evolution of small black holes in spiral galaxies,
which lead to relatively low average spins for these black holes
\citep{v07,kph08}. Motivated by their results on the cosmological
spin evolution of black holes, we tentatively adopt a $M_{\rm
bh}$-dependent radiative efficiency, in which $\eta_{\rm rad}$
remains constant for $M_{\rm bh}< 10^8$ and increases as a power-law
with black hole mass for $M_{\rm bh}\ge 10^8$: }
\begin{equation}
 \eta_{\rm rad}=\left\{ \begin{array}{ll}
        \eta_{\rm rad,0}, & \mbox{if $M_{\rm bh}< 10^8 {\rm M}_{\odot}$};\\
         \eta_{\rm rad,0}\left({\frac {M_{\rm bh}}{10^8 {\rm M}_{\odot}
}}\right)^q, & \mbox{if $M_{\rm bh}\ge 10^8 {\rm M}_{\odot}$
}.\end{array} \right. \label{etarad}
        \end{equation}
in our calculations. We find that the measured local BHMF can be
roughly reproduced by the BHMF of AGN relics provided $\eta_{\rm
rad,0}=0.08$ and $q=0.36$ are adopted (see the lower panel in Fig.
\ref{fig3}).

\begin{figure}
\centerline{\psfig{figure=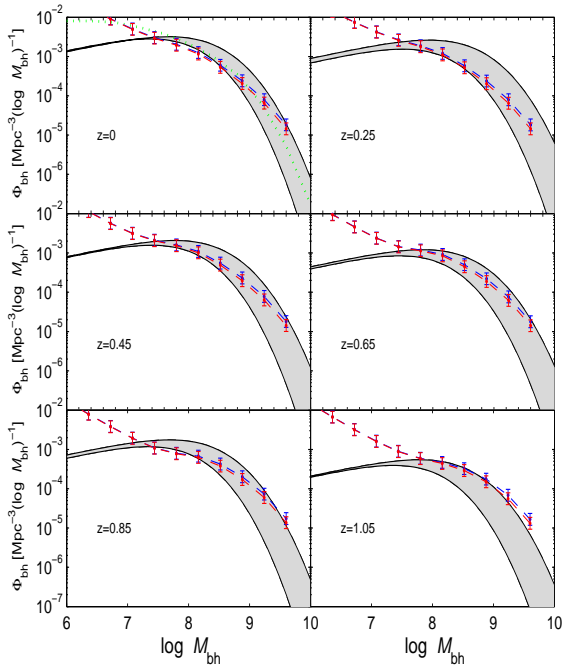,width=7.5cm,height=9.0 cm}}
\caption{The BHMFs at different redshifts of $z=0$, 0.25, 0.45,
0.65, 0.85, and 1.05  transformed from the COMBO-17 LFs are
indicated with shaded areas \citep*[taken from][]{t06}. The BHMFs of
AGN relics at different redshifts calculated from the AGN LF are
plotted by using $M_{\rm bh}$-dependent radiation efficiency
$\eta_{\rm rad}(M_{\rm bh})$ (see the text for the details). The
blue lines are calculated with a correction factor of 1.6 in
dependent of luminosity for Compton-thick AGN, while the red lines
represent the cases with a luminosity-dependent correction for the
Compton-thick AGN. } \label{fig4}
\end{figure}

\citet{t06} tried to derive the BHMFs with redashifts up to $z\sim
1$ from the spheroid LF of early-type galaxies using the correlation
between the spheroid luminosity and black hole mass (see their paper
for the details), which provide further constraints on the model
calculations for the cosmological evolution of massive black holes.
The model calculations performed with this $M_{\rm bh}$-dependent
radiative efficiency (Equation \ref{etarad}) are compared with the
spheroid-BHMFs derived by \citet{t06} in Fig. \ref{fig4}. We find
that calculated BHMFs of AGN relics can roughly reproduce the
spheroid-BHMFs at different redshifts either for
luminosity-dependent or luminosity-independent corrections for the
Compton-thick AGN.

\section{Discussion}

As in the most previous works, we implicitly assume that the black
hole growth is dominated by mass accretion in bright AGN, while some
inactive black holes may still be accreting gases, though their mass
accretion rates are very low. If the duration of the accretion in
these objects is as long as the Hubble timescale, they can accrete
sufficient mass comparable with that accumulated in bright AGN
phases, as the AGN phase is much shorter than the Hubble timescale.
It is believed that the advection dominated accretion flows (ADAFs)
are present in those objects, which are very hot and radiate mostly
in hard X-ray bands \citep{ny94}. They are very difficult to be
detected due to low luminosity, unless those in the nearby Universe.
\citet{cao05} suggested that the accretion of such low-luminosity
objects can be constrained by the hard X-ray background, though the
emission from most of these individuals cannot be detected by any
facilities now. It was found that less than $\sim 5$ per cent of the
local black hole mass density was accreted during the ADAF phases,
which will be even lower if the Compton-thick AGN are included
\citep*[see][for the details]{cao07}. \citet{h06} considered the
distribution of local supermassive black hole Eddington ratios and
accretion rates, accounting for the dependence of radiative
efficiency and bolometric corrections on the accretion rate. They
also found that black hole mass growth was dominated by AGN phase,
and not by the radiatively inefficient low accretion rate phase in
which most local supermassive black holes are currently observed.

The main difference of this work from the previous works is that the
Eddington ratio distributions are derived from an AGN sample with
measured Eddington ratios \citep{k06}. The Eddington ratio
distributions for fixed black hole mass derived in our work
approximate to the lognormal distribution (see Fig. \ref{fig1}), and
the mean Eddington ratios are in the range of $\sim 0.1-0.3$ varying
with black hole mass and redshift (see Fig. \ref{fig2}). For most
cases, the mean Eddington ratios peak at $\sim 10^8{\rm M}_\odot$,
and then decline with increasing black hole mass. Even at the low
mass end, most black holes are accreting at $\lambda\sim0.2$, which
implies that the objects accreting at extremely high rates should be
rare or such phases are very short. It was suggested that the
radiative efficiency $\eta_{\rm rad}$ declines for a slim accretion
disc provided the mass accretion rate is sufficiently high due to
the photon trapping effort \citep[e.g.,][]{a88,b78,w99}.
\citet{w00}'s calculations on the slim discs showed that the
radiative efficiency will not deviate significantly from that for
standard thin discs if $L_{\rm bol}/L_{\rm Edd}\la 2$, which implies
that the present adopted radiative efficiency independent of
Eddington ratio $\lambda$ is indeed a good assumption.

There is only one free parameter $\eta_{\rm rad}$ in our
calculations for the cosmological evolution of massive black holes.
We find that the resulted BHMF of AGN relics is unable to reproduce
the measured local BHMF for any value of $\eta_{\rm rad}$ adopted,
provided the radiative efficiency $\eta_{\rm rad}$ is independent of
black hole mass, as treated in previous works
\citep*[e.g.,][]{yt02,m04,t06}. The mean Eddington ratios adopted in
our calculations are derived from an AGN sample, which are in the
range of $\sim 0.1-0.3$. Thus, it is not surprising that the local
BHMFs cannot be reproduced by our calculations with any constant
radiative efficiency, because the Eddington ratio $\lambda\sim 1$ is
usually required in order to let the resulted BHMF match the local
one in those works. In this work, we use two different corrections
(either luminosity-independent or luminosity-dependent) for the
Compton-thick AGN (see Sect. 3 for the details), and find that the
final results are quite similar (see Figs. \ref{fig3} and
\ref{fig4}). We also use the hard X-ray LF derived from an AGN
sample at high redshifts by \citet{s08} in stead of the bolometric
LF of \citet{h07} in the calculations. It is found that the main
results of this work change very little and the main conclusion is
not altered.

\citet{v07} studied on how the accretion from a warped disc
influences the evolution of black hole spins and concluded that
within the cosmological framework, one indeed expects most
supermassive black holes in elliptical galaxies to have on average
higher spin than black holes in spiral galaxies, where random, small
accretion episodes (e.g., tidally disrupted stars, accretion of
molecular clouds) might have played a more important role. Thus, we
tentatively adopt a $M_{\rm bh}$-dependent radiative efficiency (see
Eq. \ref{etarad}), in which $\eta_{\rm rad}$ remains constant for
$M_{\rm bh}\le 10^8$ and increases with black hole mass for $M_{\rm
bh}> 10^8$. This $M_{\rm bh}$-dependent radiative efficiency is
qualitatively consistent with the results of \citet{v07}. It is
found that the measured BHMFs can be fairly well reproduced by our
model calculations with this $M_{\rm bh}$-dependent radiative
efficiency (see Figs. \ref{fig3} and \ref{fig4}), which require
$\eta_{\rm rad}\ga 0.18$ for $M_{\rm bh}\ga 10^9{\rm M}_\odot$. This
provides evidence for most massive black holes being spinning very
rapidly. It is interesting to find that $a\simeq 0.9$ is also
required by the fitting of the residual hard X-ray background with
the emission from the ADAFs in the low-luminosity objects
\citep{cao07}.
Our calculations can be improved if
the mean spin parameter $a$ as a function of black hole mass is
available from the work within the cosmological framework, which is
beyond the scope of present work.

{  In our present calculations of the black hole evolution, the
black hole mergers have been neglected. \citet{s07} assessed the
importance of the black hole mergers on the evolution of the BHMF.
They found that the impact of black hole mergers on the cosmological
evolution of BHMF may probably be small compared with black hole
accretion processes, while its impact on the black hole spin
evolution may be important \citep*[e.g.,][]{wc95,hb03,v05,v07}. The
effect of black hole mergers increases the number density of very
massive black holes \citep{s07}, which implies that the radiative
efficiencies for very massive active black holes should be higher
than the present values if black hole mergers are included in our
calculations. This strengthens our conclusion that most massive
black holes are spinning very rapidly. }

\section*{Acknowledgments}
We thank the anonymous referee for the helpful comments/suggestions,
T. G. Wang and Y. F. Yuan for discussion, and N. Tamura for
providing us the data of the spheroid-BHMFs. This work is supported
by the NSFC (grant 10773020), and the CAS (grant KJCX2-YW-T03).

\end{document}